# The Heaviside equation for laser heating of the fullerennes

Janina Marciak-Kozlowska[1]
Miroslaw Kozlowski[2]

[1] Institute of Electron Technology, Al. Lotnikow 32/46, 02 – 669 Warsaw, Poland
[2] Corresponding author, e-mail: miroslawkozlowski@aster.pl

Abstract

In his paper the heating of the fullerenes by ultra-short laser pulses is investigated. The thermal Heaviside equation is formulated and solved for the Cauchy initial condition The effective thermal relaxation time is calculated..

Introduction

In order to find out why the quantum mechanics is limited to the world of photons and elementary particles the decoherence processes must be investigated. For macroscopic particles there are two main "natural" way of decoherence. On the one hand collisions with other particles, and on the other hand the thermal emission of radiation due to the internal heat of an object. The warm object, i.e. macromolecules, emits the photons which can be detected, and the decoherence can be measured.

Of the macromolecules fullerenes have the interesting feature that they can be turned into a cross border commuter between the classical and quantum description. When they are relatively cold (900K) they can show perfect quantum behavior, but they posses already sufficiently many degrees of freedom to store and partially release thermal energy in form of photons. The amount of thermal energy stored can be controlled by measurement of their temperature [1,2].

In this paper we investigate in detail the thermal processes in fullerenes excited by ultra-short laser pulses. Based on the results presented in monograph [3] we develop the quantum heat transport equation for the thermal energy propagation in fullerenes. The Heaviside quantum equation is solved for the Cauchy initial condition. It is shown that the temperature field $T(x,t)$ of the fullerenes heated by ultra-short laser pulses consists of two components: ballistic and diffused. Only the ballistic component contains the information on the initial of the heating process.

## 2. The model

The most prominent example of the quantum materials are the fullerenes. Fullerenes have bees the subject of many fascinating fundamental dynamical studies in the field of atomic and molecular physics. They are an allotrope of pure carbon where the carbon atoms form a closed hollow cage. The most famous example of the fullerenes is Buckminster fullerene, $C_{60}$. The molecule has the geometrical form of a truncated dodecahedron with a carbon atom sitting at each corner of the polyhedron.

In this paper we propose the model for the thermal relaxation phenomena in the fullerenes in the frame of quantum heat transport equation formulated in monograph [3]. In the following starting with elementary relaxation time $\tau^e$ and $\upsilon^e$, viz:

$$\tau^e = \frac{\hbar}{m\upsilon_e^2}, \qquad \upsilon^e = \alpha c$$

we will describe thermal relaxation process in fullerenes which consist of $N$ atoms (i.e. $C^{60}$) each with elementary $\tau^e$ and $\upsilon^e$. To that aim we use the Pauli – Heisenberg inequality [1]

$$\Delta r \Delta p \geq N^{\frac{1}{3}} \hbar \qquad (1)$$

The Pauli – Heisenberg inequality expresses the basic property of the $N$ – fermionic system. In fact, compared to the standard Heisenberg inequality

$$\Delta r \Delta p \geq \hbar \qquad (2)$$

we notice that in this case the presence of the larger number of identical fermions forces the system either to become spatially more extended for fixed typical momentum dispersion, or to increase its typical momentum dispersion for a fixed typical spatial extension. We could also say that for a fermionic system in its ground state, the average energy per particle increases with the density of the system.

A picturesque way of interpreting the Pauli – Heisenberg inequality is to compare Eq.(1) with Eq.(2) and to think of the quantity on the right side of it as "the effective" fermionic Planck constant

$$\hbar^{eff}(N) = N^{\frac{1}{3}} \hbar \qquad (3)$$

According to formula (3) we calculate the effective relaxation time $\tau^{eff}(N)$ and effective thermal velocity $\upsilon^{eff}(N)$

$$\upsilon^{eff}(N) = \frac{1}{N^{\frac{1}{3}}}\upsilon^e \tag{4}$$

$$\tau^{eff}(N) = N\tau^e \tag{5}$$

In monograph [1] the Heaviside quantum thermal heat transport was formulated

$$\frac{1}{\upsilon^2}\frac{\partial^2 T}{\partial t^2} + \frac{m}{\hbar}\frac{\partial T}{\partial t} = \frac{\partial^2 T}{\partial x^2} \tag{6}$$

Considering Eqs. (4,5), transport equation (6) can be written as:

$$\frac{1}{(\upsilon^{eff})^2}\frac{\partial^2 T^{eff}}{\partial t^2} + \frac{m}{\hbar^{eff}}\frac{\partial T^{eff}}{\partial t} = \frac{\partial^2 T^{eff}}{\partial x^2} \tag{7}$$

Let us consider the solution of the Eq.(7) for the Cauchy initial condition

$$T^{eff}(x,0) = 0, \quad T^{eff}(0,t) = f(t) \tag{8}$$

For initial condition (8) the solution of Eq.(8) has the form [3]:

$$T^{eff}(x,t) = \left\{ f\left(t - \frac{x}{\upsilon^{eff}}\right)\exp\left(-\frac{\rho x}{\upsilon^{eff}}\right) + \frac{\sigma^{eff} x}{\upsilon^{eff}}\int_{x/\upsilon^{eff}}^{t} f(x-y)\exp(-y\rho)\frac{I_1\left[\sigma\left(y^2 - \frac{x^2}{(\upsilon^{eff})^2}\right)^{1/2}\right]}{\left(y^2 - \frac{x^2}{(\upsilon^{eff})^2}\right)^{1/2}}dy \right\} \times$$
$$\times H\left(t - \frac{x}{\upsilon^{eff}}\right) \tag{9}$$

where $H\left(t - \frac{x}{\upsilon^{eff}}\right)$ is the Heaviside function.

In formula (9)

$$\upsilon^{eff} = \frac{e^2}{\hbar^{eff}}$$

$$\tau^{eff} = \frac{\hbar^{eff}}{m(\upsilon^{eff})^2} \tag{10}$$

$$\rho^{eff} = \sigma^{eff} = \frac{1}{2\tau^{eff}}$$

as can be seen from formula (9) the effective temperature field $T^{eff}(x,t)$ of the fullerenes has the two components $T_B^{eff}(x,t)$ – ballistic and $T_D^{eff}(x,t)$ – diffusion, i.e.

$$T_B^{eff}(x,t) = f\left(t - \frac{x}{v^{eff}}\right) \exp\left(-\frac{\rho^{eff} x}{v^{eff}}\right) H\left(t - \frac{x}{v^{eff}}\right)$$

$$T_D^{eff}(x,t) = \left\{\frac{\sigma^{eff} x}{v^{eff}} \int_{x/v^{eff}}^{t} f\left(x - \frac{x}{v^{eff}}\right) \exp(-y\rho^{eff}) \frac{I_1\left[\sigma^{eff}\left(y^2 - \frac{x^2}{(v^{eff})^2}\right)^{1/2}\right]}{\left(y^2 - \frac{x^2}{(v^{eff})^2}\right)^{1/2}} dy\right\} H\left(t - \frac{x}{v^{eff}}\right) \quad (11)$$

For times of the order of the effective relaxation time $\tau^{eff}$ (ballistic pulse) the heat pulse preserves its shape and only the amplitude is diminished due to the scattering. On the other hand for the longer time periods a new structure develops. Multiple scattering distort the shapes of the initial pulse. One can say that for $t \gg \tau^{eff}$ the information contained in the initial pulse is lost as the time $t \to \infty$. When fullerenes are heated with ultra short laser pulses the two steps should be observed. For the time shorter than relaxation time fullereness are "cold" and only for times longer than relaxation time they are "hot"

3. Conclusions

In this paper the thermal heating of the fullerenes by ultra-short laser pulses is investigated. The thermal transport equation is formulated and solved for the fullerenes with $N$ atoms. It is shown that the effective thermal relaxation time is scaled as $N$.

References

[1]     L. Hackermüller et al., *Nature* (2004) 711
[2]     L. Hackemüller et al., *Phys. Rev. Lett.* **91** (2003) 090408

[3] M. Kozlowski, J. Marciak – Kozlowska, *From quarks to bulk matter,* Hadronic Press, USA, 2001.